# Mechanisms driving robust high-temperature superconductivity in complex metal hydrides under moderate pressure

Wendi Zhao[1], Shumin Guo[1], Tiancheng Ma[1], Zhengtao Liu[1], Chengda Li[1], Defang Duan[1*], and Tian Cui[1,2*]

[1]State Key Laboratory of Superhard Materials, College of Physics, Jilin University, Changchun 130012, China

[2]Institute of High Pressure Physics, School of Physical Science and Technology, Ningbo University, Ningbo 315211, China

*Corresponding authors. duandf@jlu.edu.cn (D. D), cuitian@nbu.edu.cn (T. C)

**ABSTRACT**

The discovery of near-room-temperature superconductivity in compressed hydrides has sparked intensive research efforts to identify superconducting hydrides stable at low or even ambient pressures. Herein, we demonstrate a new mechanism for achieving robust superconductivity in complex metal hydrides under moderate pressure, using $Li_3IrH_9$ as a paradigmatic example. This compound displays unique electronic structural characteristics where the broadening and overlap between antibonding electronic bands of $[IrH_8]^{2-}$ and adjacent $H^-$ orbitals not only drive the intrinsic metallicity of the hydrogen sublattice, generating hydrogen-dominated electronic states at the Fermi level, but also soften hydrogen-related optical phonon modes, inducing strong electron-phonon coupling that remains robust even under high-pressures. First-principles calculations predict that $Li_3IrH_9$ maintains thermodynamic stability at 100 GPa while exhibiting a consistently high $T_c$ exceeding 100 K across a broad pressure range (8-150 GPa). Through high-throughput computational screening, we have identified a new superconducting family based on this structural prototype, including $Li_3RhH_9$ ($T_c$ = 124 K at 20 GPa) and $Li_3CoH_9$ ($T_c$ = 80 K at 10 GPa). This work provides a new platform and original theoretical insights for the development of complex metal hydride superconductors that exhibit robust high-temperature superconductivity and promising practical applications.

## INTRODUCTION

Compressed hydrides have emerged as forefront candidates for high-temperature superconductivity, with binary systems such as $H_3S$[1-3] and $LaH_{10}$[4-9] and $YH_6$[10, 11] demonstrating critical temperatures ($T_c$) exceeding 200 K. Theoretical studies further propose that ternary clathrate hydrides (e.g., $Li_2MgH_{16}$[12], $Li_2NaH_{16}$, $Li_2NaH_{17}$[13], $LaSc_2H_{24}$[14]) could achieve room-temperature superconductivity[14-16]. Notably, the realization of these remarkable superconductivity invariably requires extreme stabilization pressures, presenting substantial challenges for practical applications, thereby motivating intensive research to stabilize superconducting phases under moderate or ambient pressures.

The modulation of hydrogen-bonding networks critically governs pressure requirements in superconducting hydrides[15]. Ternary clathrates such as $YCaH_{12}$[16] and $YLuH_{12}$[17] maintain high stabilization pressures comparable to binary systems owing to preserved H-H bonding motifs, while alloy-based hydride systems achieve significant pressure reduction through structural reorganization—notably maintaining exceptional stability even at substantially lower pressures. For example, $LaBeH_8$ adopts a fluorite-type structure derived from clathrate $LaH_{10}$, stabilizing below megabar pressures with a measured $T_c$ of 110 K at 80 GPa [18, 19]. Unlike the pure hydrogen cage lattice in $LaH_{10}$, the Be-H alloy lattice pre-compressed by La element can be can be sustained at significantly lower pressure. Similar mechanisms operate in $LaBH_8$ (154 K at 50 GPa), $BaSiH_8$ (71 K at 3 GPa), and $SrSiH_8$ (126 K at 27 GPa), ensuring their dynamic stability under moderate pressures[20-22]. Notably, the $KB_2H_8$ system derived from $LaH_{10}$ through elemental substitution exhibits a superconducting transition up to 146 K at 12 GPa[23], featuring $BH_4$ tetrahedral molecules embedded in a face-centered cubic potassium lattice, with thermal stability achieved around 50 GPa[22]. Analogous structures including $RbB_2H_8$ (101 K at 15 GPa)[24], $MgC_2H_8$ (55 K at 40 GPa)[25] and $AlN_2H_8$ (118 K at 40 GPa)[26] are predicted to exhibit excellent superconductivity under moderate pressures.

In contrast to high-pressure superhydrides, complex transition metal hydrides

(CTMHs), as a unique class of multifunctional hydrogen-rich materials, have attracted considerable interest in hydrogen storage and catalytic applications[27]. These compounds feature negatively charged homoleptic hydrido complexes $[TMH_n]^{m-}$, where transition metals (TM) form covalent bonds with multiple H atoms exhibiting diverse coordination modes, and some CTMHs also incorporate interstitial H atoms exclusively coordinated by metal cations. These complexes are stabilized via charge transfer from electropositive counterions (e.g., alkali, alkaline earth, or rare earth metals), as exemplified by experimentally synthesized systems including $K_2ReH_9$[28], $Mg_2FeH_6$[29], $Na_3WH_9$ [30], $Mg_3CrH_8$ [31], $BaReH_9$[32], and $Li_5MoH_{11}$[33, 34]. While these materials have been extensively explored for energy-related applications[27], their superconducting potential remains largely overlooked owing to their wide-bandgap insulating nature. Current experimental reports have only identified weak superconductivity in pressurized $BaReH_9$ ($T_c$ = 7 K at 102 GPa) [32] and $Li_5MoH_{11}$ ($T_c$ = 6.5 K at 160 GPa) [33]. Notably, recent theoretical studies predict that $Mg_2IrH_6$—a thermodynamically metastable CTMH—could host an extraordinary $T_c$ of 120~160 K under ambient pressure [35-37]. However, the direct synthesis of this metastable phase under ambient conditions remains experimentally challenging. A promising development emerged with the successful synthesis of $Mg_2IrH_5$[38], which theoretically can serve as a precursor for accessing the superconducting $Mg_2IrH_6$ phase. These findings underscore the considerable, yet largely untapped, potential for high-$T_c$ superconductivity within CTMH systems. However, fundamental challenges persist—particularly in understanding the underlying mechanisms governing material both stability and superconducting behavior — which hinder systematic exploration of superconductivity across this important class of complex hydrides.

Inspired by the excellent superconductivity of $Mg_2IrH_6$[36], we performed a large-scale structural prediction of the ternary Li-Ir-H system, leading to the discovery of an unprecedented complex transition metal hydride, $Li_3IrH_9$. This compound features anionic $[IrH_8]^{2-}$ units and an interstitial $H^-$ ion encapsulated within Li octahedra, which is thermodynamically stable above 100 GPa, and dynamically stable down to 8 GPa with $T_c$ up to 113 K. The exceptional superconducting properties originate from a distinctive electronic mechanism: the broadening and overlap of the antibonding electron bands from the $[IrH_8]^{2-}$ units with the interstitial $H^-$-derived bands drive the intrinsic metallicity of the hydrogen sublattice. Notably, the metallic interstitial $H^-$ ions contribute dominantly to electron-phonon coupling, retaining robustness even high pressure. The predicted $Li_3IrH_9$ offers an unprecedented hydride structural prototype exhibiting robust high-temperature superconductivity, particularly considering that its isotypic counterparts have shown excellent superconducting properties at moderate pressures.

**RESULTS AND DISCUSSION**

We performed large-scale random structure prediction for the Li-Ir-H system at 50 GPa and 100 GPa using the AIRSS code [39, 40], generating over 30,000 candidate structures. To construct the ternary phase diagram for assessing thermodynamic stability, we reproduced previously reported stable binary hydrides [41, 42] and elemental solids[43-45] at the relevant pressures and discovered three novel stable alloys in the Li-Ir system: $Li_3Ir$, $Li_2Ir$, and $Li_2Ir_3$. The ternary phase diagram at 100 GPa revealed seven thermodynamically stable ternary Li-Ir-H compounds, including: *Pm*-3*m* $LiIr_3H_4$, *P*-62*m* $LiIr_3H_9$, *P*4/*mmm* $LiIrH_2$, *C*2/*c* $LiIrH_4$, *I*4/*mmm* $Li_3IrH_4$, *I*4$_1$/*amd* $Li_2IrH_7$, and the particularly noteworthy *Pm*-3*m* $Li_3IrH_9$ (see Figure S1-2). We further evaluated their dynamic stability threshold pressures and corresponding $T_c$ values. All thermodynamically stable phases maintain dynamic stability at moderate pressures (see Table S1). Notably, *Pm*-3*m* $LiIr_3H_9$ remains dynamically stable at as low as 8 GPa and exhibits a remarkably high $T_c$ of 113 K, significantly exceeding the maximum $T_c$ values of the other stable phases. Strikingly, even at 100 GPa, $LiIr_3H_9$ retains a high $T_c$ of 115 K, demonstrating robust superconductivity. Given the goal of searching for high $T_c$ hydrides at moderate pressures, $Li_3IrH_9$, which has excellent stability and superconductivity, will be the focus of our next investigations, aiming to elucidate the fundamental structure-property relationships governing this class of metallic hydrides.

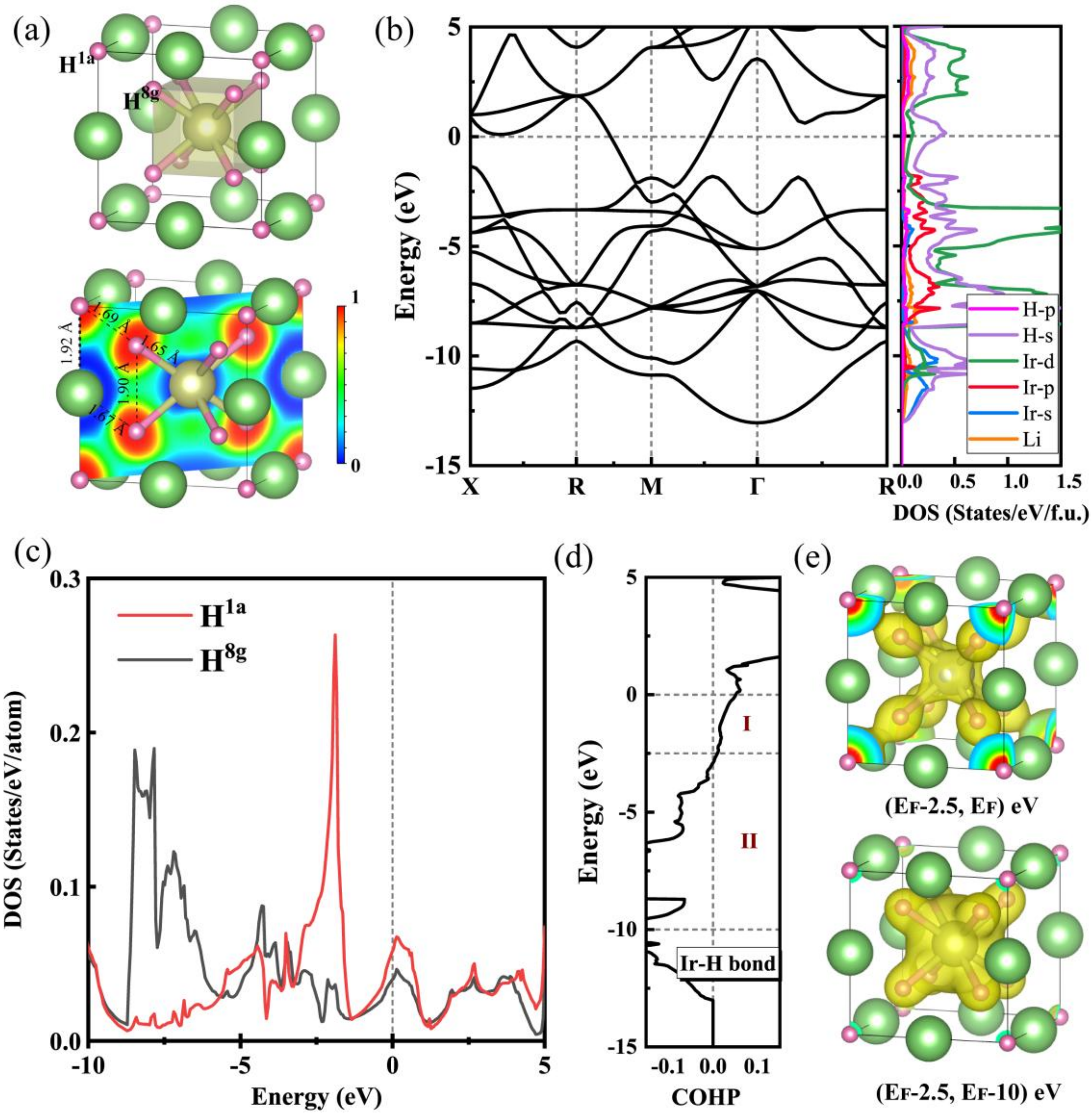

**Figure 1.** *Pm*-3*m* $Li_3IrH_9$ at 50 GPa. (a) Crystal structure and electron localization function (ELF). (b) Electronic band structure and density of states (DOS). (c) Projected DOS of each $H^{1a}$ and $H^{8g}$ atom. (d) Crystal orbital Hamilton population (COHP) of the Ir-H bond. (e) Calculated partial charge density in the specified energy ranges ($E_F$-2.5, $E_F$) and ($E_F$-2.5, $E_F$-10), corresponding to partitions I and II of the COHP, respectively.

The crystal structure of *Pm*-3*m* $Li_3IrH_9$ is illustrated in Figure 1a. In this structure, Li and Ir atoms occupy the 3d and 1b Wyckoff positions, respectively. The H atom (labeled $H^{1a}$), located at the 1a Wyckoff position, is encapsulated at the center of the Li octahedron, which can be viewed as a fragment of the *Fm*-3*m* LiH lattice. Meanwhile, other H atoms (labeled $H^{8g}$) occupying the 8g Wyckoff positions form $IrH_8$ cubic unit around each Ir atom. At 50 GPa, the nearest-neighbor Li-H and Ir-H distances are 1.67

Å and 1.65 Å, respectively. The nearest-neighbor H-H distance is 1.69 Å, significantly longer than the bond length of the $H_2$ molecule (~0.74 Å) and even exceeding the typical H-H distances observed in clathrate hydride superconductors (1.1-1.3 Å)[46]. Metal atoms exhibit low electronegativity, with alkali metals serving as excellent electron donors. Bader charge analysis reveals that each Li and Ir atom donates electrons 0.83 |e| and 0.24 |e| to H atoms, respectively. Coordinating with more Li atoms than $H^{8g}$, the $H^{1a}$ atom acquires 0.46 |e|, higher than the 0.29 |e| obtained by each $H^{8g}$ atom. This charge distribution is further corroborated by electron localization function (ELF) analysis, which demonstrates purely ionic bonding between Li and H atoms, evidenced by the absence of electron localization in these regions. In contrast, the Ir-H interactions exhibit weak electron localization, indicative of polarized covalent bonding that characterizes the $IrH_8$ molecular units. Collectively, these observations collectively define the fundamental building blocks of $Li_3IrH_9$ as consisting of covalently bonded $[IrH_8]^{2-}$ unit and interstitial $H^-$ ion, stabilized through charge transfer from Li to form an ionic configuration represented as $Li_3^+[IrH_8]^{2-}H^-$. Interestingly, despite the relatively large interatomic distances, weak electron localization persists between hydrogen atoms, particularly between $H^{1a}$ and $H^{8g}$, where the ELF value at their midpoint reaches approximately 0.37. This suggests a weak covalent interaction within the hydrogen sublattice and confirms its the intrinsic metallicity, distinguishing it markedly from the wide bandgap insulating behavior of typical conventional CTMHs.

Figure 1b presents the electronic band structure and density of states (DOS) of $Li_3IrH_9$ at 50 GPa. The localized DOS peak corresponding to Ir-*d* orbitals lies well below the Fermi energy, while hydrogen-derived states dominate the Fermi-level region, exhibiting a pronounced Van Hove singularity. Notably, the contributions of H atoms occupying different Wyckoff positions to the DOS near the Fermi level are significantly distinct. Each $H^{1a}$ atom exhibits higher DOS contributions at the Fermi level compared to $H^{8g}$, with a distinct DOS peak adjacent to the Fermi level (Figure 1c). In contrast, the primary DOS peaks of $H^{8g}$ atoms are localized between -10 eV and -5 eV, which is attributed to the strong hybridization between the Ir-d and H-s states in this region. This can be revealed more rigorously by atomic orbital symmetry analysis, that is, $H^{8g}$ exhibits greater orbital symmetry matching with Ir atoms compared to $H^{1a}$, implying strong coupling (see Table S2). Additionally, weak coupling in the $H^{1a}$-$H^{8g}$ interaction arises from s-orbital hybridization, which serves as the key driver for the intrinsic metallicity of the hydrogen sublattice.

For typical CTMHs, the transition metals tend to form homoleptic complex ions with a high number of hydrogen atoms (e.g., $[ReH_9]^{2-}$[32], $[MoH_9]^{3-}$[33], $[IrH_6]^{3-}$[47, 48], $[IrH_5]^{4-}$[49]), where the bonding mechanism usually follows the 18-electron rule. The substantial hybridization between transition metal and hydrogen orbitals under ligand field effects promotes the formation of strongly occupied $\sigma$-bonding states, while the antibonding orbitals remain unoccupied. This electronic configuration induces a wide band gap near the Fermi level, suppressing electron delocalization. For example, in ambient-pressure $BaReH_9$, the isolated $[ReH_9]^{2-}$ units exhibit no electron localization between neighboring complexes and maintain purely ionic interactions with $Ba^{2+}$ ions, preventing the formation of H-H covalent networks (see Figure S4). Pressure generally enhances orbital overlap between homoleptic complex ions, promoting electron delocalization and metallization. $BaReH_9$ is experimentally measured to enter the superconducting phase above 100 GPa[32].

In contrast to conventional CTMHs, $Li_3IrH_9$ exhibits a distinct electronic configuration where extra electrons in the $[IrH_8]^{2-}$ anionic units occupy antibonding orbitals ($A_{1g}^*$), which are further delocalized through the crystal periodicity in $Li_3IrH_9$, forming highly extended electronic states (see Figure S5). Similar diffuse antibonding orbitals for Ir-H bonds also appear in $Mg_2IrH_6$ and play an important role in inducing superconductivity[50]. The calculated COHP confirms that the Ir-$H^{8g}$ bond exhibits pronounced antibonding characteristics near the Fermi level. Crucially, these diffuse antibonding orbitals extend along the direction of the Ir-$H^{8g}$ bond toward the $H^{1a}$ site, thereby further enhancing electron delocalization (Figure 1d-e)[51, 52]. The crystal orbitals occupying deeper energy levels are predominantly localized within the $IrH_8$ units, displaying significant bonding characteristics that further substantiate the covalent nature of Ir-$H^{8g}$ interactions. Hence, the intrinsic metallicity of hydrogen sublattices is primarily driven by the broadening and overlapping between the electronic bands of $H^{1a}$ ions and Ir-$H^{8g}$ antibonding states. This behavior mimics the electronic behavior of atomic metallic hydrogen[53], thereby facilitating phonon-mediated superconductivity. Meanwhile, strong ionic interactions between the highly electropositive counterions $Li^{+}$ and the anionic $[IrH_8]^{2-}$ units and $H^{-}$ ions play an important role in stabilizing the lattice. It is worth emphasizing that while some CTMHs also contain interstitial $H^{-}$ which is exclusively coordinated by metal cations, these typically do not contribute to superconductivity. This arises primarily from the absence of antibonding electron occupation in the complex anions $[TMH_n]^{m-}$ combined with lower structural symmetry that diminishes orbital overlap between $H^{-}$ ions and the $[TMH_n]^{m-}$ units. For instance, in $Mg_3CrH_8$, the isolated interstitial $H^{-}$ ions primarily improve structural stability through charge-transfer-enhanced ionic interactions with active metal cations (see Figure S4). These findings provide crucial insights into the cooperative roles of complex anions $[TMH_n]^{m-}$ and interstitial $H^{-}$ in enabling high-temperature superconductivity in CTMHs.

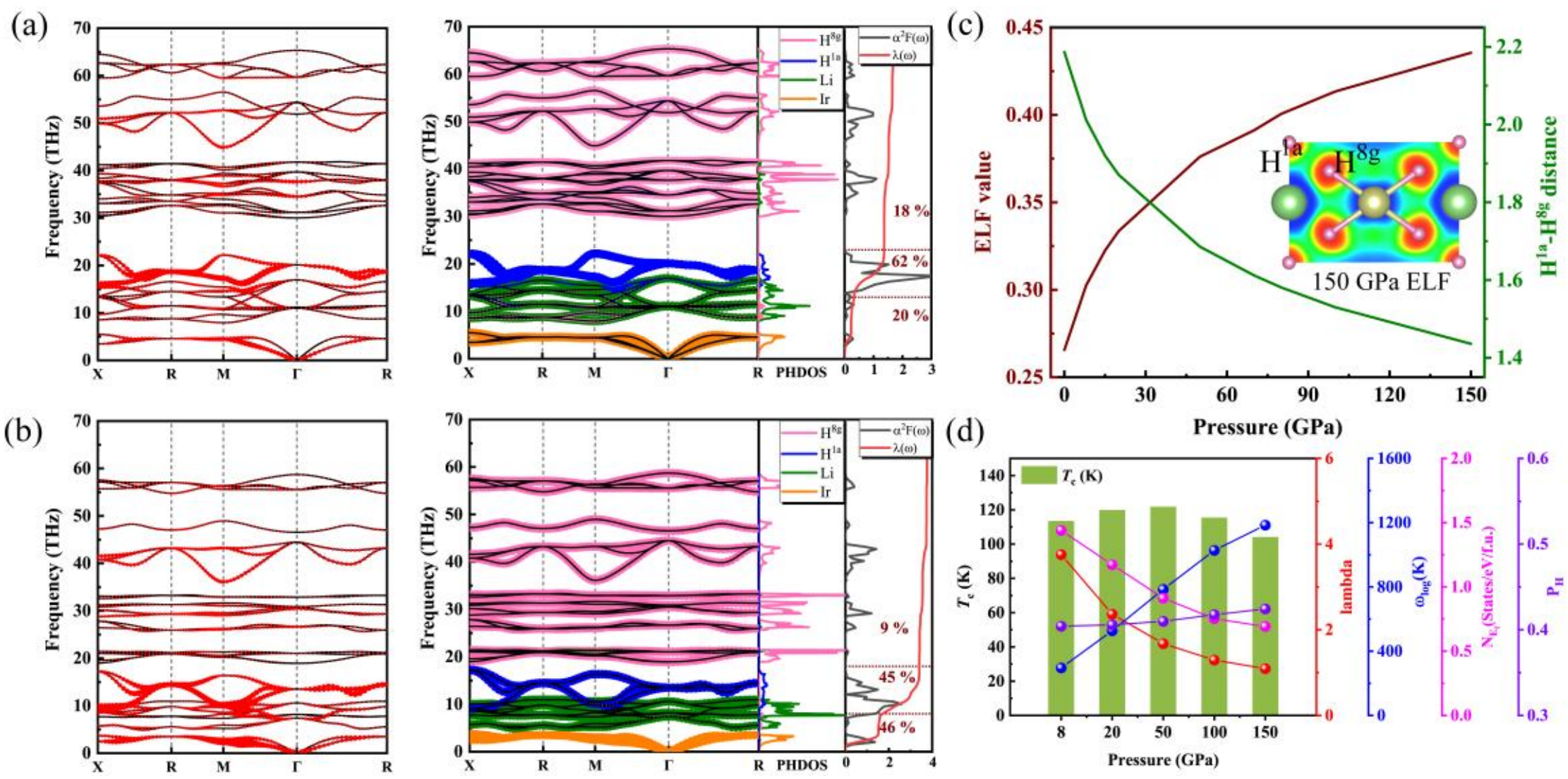

**Figure 2** Phonon dispersion curves, projected phonon density of states (PHDOS), and Eliashberg spectral function $\alpha^2F(\omega)$ along with the electron-phonon integral $\lambda(\omega)$ for $Li_3IrH_9$ at (a) 50 GPa and (b) 8 GPa. The size of the red dots on the phonon dispersion curves represents the contribution of different phonon modes to the EPC,

while the atomic-resolved phonon modes are also illustrated. (c) Pressure dependence of the $H^{1a}$-$H^{8g}$ distance in $Li_3IrH_9$ and its ELF value at the bond center. (d) Calculated superconducting parameters of $Li_3IrH_9$ at different pressures. The obtained $T_c$ using Eliashberg equation with $\mu^* = 0.1$[54], the EPC parameter ($\lambda$), logarithmic average phonon frequency $\omega_{log}$, the electronic density of states at the Fermi level $N(\varepsilon_F)$, the contribution of H atoms DOS to the total DOS at the Fermi energy ($P_H$).

Figure 2a-b illustrates the phonon dispersion, projected phonon density of states (PHDOS), and Eliashberg spectral functions together with the electron-phonon integral $\lambda(\omega)$ of $Li_3IrH_9$ at 50 GPa and 8 GPa, providing insights into the origin of its high-temperature superconductivity. The phonon spectrum shows a clear division of vibrational contributions: the acoustic modes are dominated by Ir atoms; the low-frequency optical modes primarily involve Li vibrations; the intermediate and high-frequency optical phonon modes are almost exclusively associated with hydrogen vibrations. The size of the red dots decorating the specific phonon modes represent their contribution to the EPC, with the largest contributions concentrated in the optical branches within the frequency range of 13–23 THz, accounting for about 62 % of the total EPC. This is further evidenced by the sharp peaks in the Eliashberg spectral function within this range. Atom-resolved phonon dispersion curves reveal that these modes are predominantly associated the vibrations of the $H^{1a}$ atom. This observation stands in sharp contrast to the vibrational properties of *Fm*-3*m* LiH at equivalent pressures: despite hydrogen atoms being similarly positioned at the center of octahedral coordination by Li atoms, their phonon frequencies are significantly higher (~40 THz) (see Figure S6-7). This difference originates from the $H^{1a}$ atoms in $Li_3IrH_9$ acquiring fewer electrons than the hydrogen atoms in LiH, combined with a longer Li-$H^{1a}$ bond length. These factors weaken the strength of the Li-$H^{1a}$ ionic bond, thereby reducing the phonon frequencies. More importantly, the weak H-H interaction in $Li_3IrH_9$ influence the vibrational behavior of the hydrogen sublattice, which is significantly different from the high-frequency vibration of isolated H ions in LiH. Although increasing pressure reduces interatomic distances in $Li_3IrH_9$ and increases ELF values

between $H^{1a}$ and $H^{8g}$ atoms as bond lengths shorten, these ELF values remain relatively weak (ELF <0.5). This indicates that weak interactions between hydrogen atoms persist throughout a wide pressure range, forming the fundamental basis for the robust superconductivity of the hydrogen sublattice (see Figure 2c).

We further investigated the pressure dependence of phonon vibration frequency. As pressure decreases, the phonon vibration frequency decreases, especially at 8 GPa, where significant softening of optical branches and even acoustic branches enhances electron-phonon coupling. At this pressure, the phonon modes of the $H^{1a}$ atoms and the Ir-dominated acoustic branch contribute almost equally to the total EPC. Interestingly, the $T_c$ of $Li_3IrH_9$ exhibits minimal pressure dependence, with values range from 104 K to 122 K over the pressure range of 8-150 GPa (Figure 2d). This robust superconductivity contrasts with the general trend in hydride superconductors, where the $T_c$ value decreases significantly with the increasing pressure. This is primarily attributed to the fact that the phonon modes of the $H^{1a}$ atom consistently maintain a dominant contribution to electron-phonon coupling under varying pressure conditions. At the same time, we examined the pressure dependence of key superconducting parameters, including $\lambda$, $\omega_{\log}$, and $N_{E_f}$, which are generally positively correlated with $T_c$. Although increasing pressure leads to a reduction in both $\lambda$ and $N_{E_f}$, the rise in $\omega_{\log}$ is more pronounced. Moreover, the almost constant high ratio of PDOS_H at the Fermi level at different pressures suggests that the hydrogen vibrational modes remain the dominant factor governing superconductivity.

For complex transition-metal hydrides (CTMHs), experimental synthesis typically involves the reaction of hydrides from active metals (alkali, alkaline-earth, or rare-earth metals) with transition metals and hydrogen gas under high-temperature and high-pressure conditions[30, 31]. Calculated formation enthalpies reveal that $Li_3IrH_9$ exhibits significant energetic advantages relative to multiple possible synthesis pathways (see Figure S11). Notably, $Li_3IrH_6$ has been successfully synthesized via the reaction of alkali metal hydrides with transition metal powders in hydrogen atmosphere[47], making it a promising precursor for $Li_3IrH_9$ synthesis. We propose three promising precursor combinations for $Li_3IrH_9$: LiH + Ir + $H_2$, $Li_3Ir$ + $H_2$, and $Li_3IrH_6$ + $H_2$, all of which exhibit favorable formation enthalpies. These results demonstrate $Li_3IrH_9$ strong potential for high-pressure synthesis and its ability to retain high-temperature superconductivity at lower pressures due to exceptional dynamic stability.

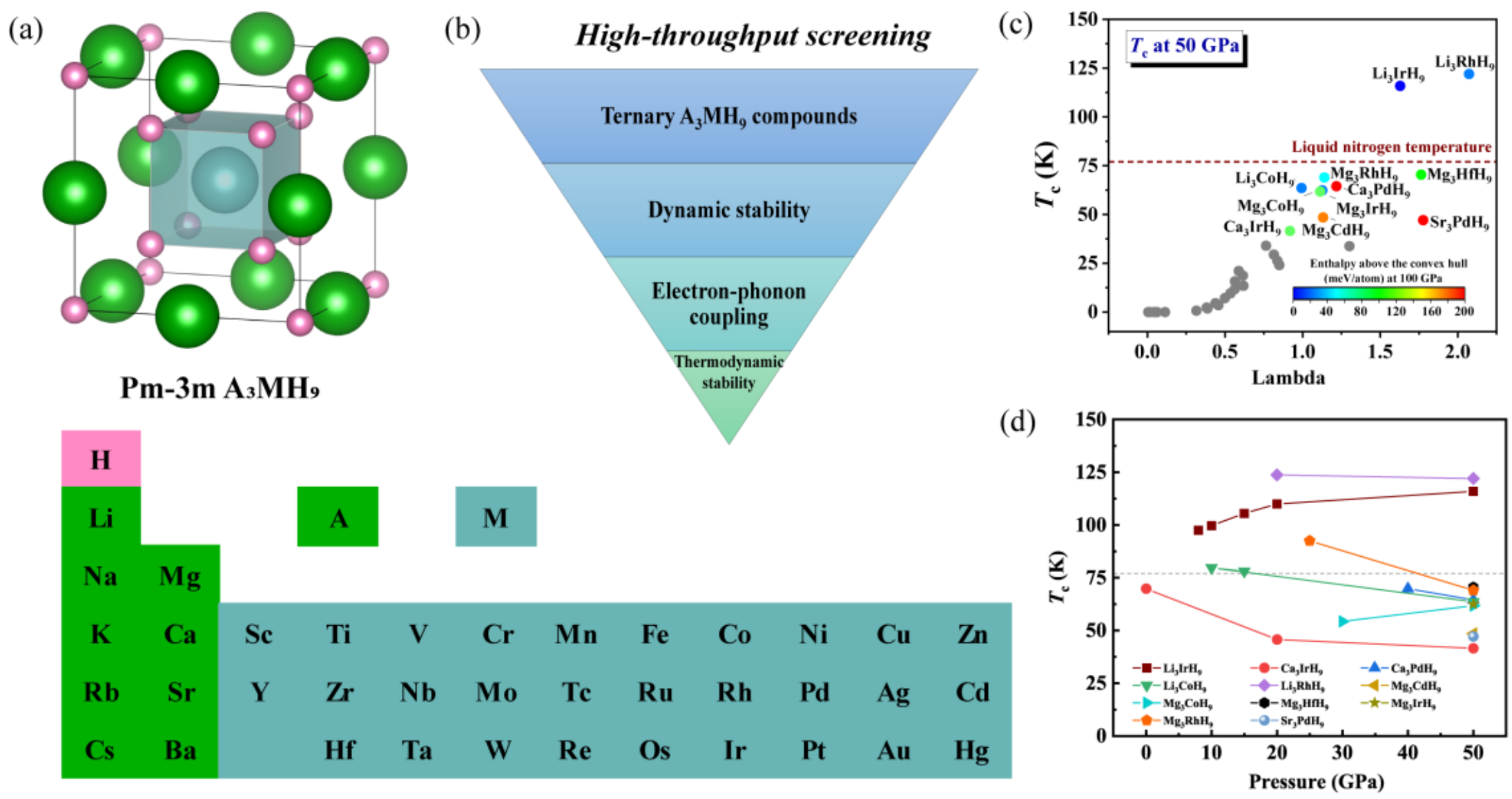

**Figure 3** (a) Crystal structure of $A_3MH_9$, (b) Procedure for high-throughput screening of stable $A_3MH_9$ compounds. Elements considered for the A and M sites, respectively, with alkali/alkaline earth metal substitutions at the A site and transition metal substitutions at the B site. (c) $A_3MH_9$ hydrides at 50 GPa. Hydrides $T_c$ higher than 40 K are labeled, with the color scale representing their thermodynamic stability (distance to the convex hull) at 100 GPa. The hydrides with $T_c$ lower than 40 K are indicated by gray points by grey points. (d) Pressure-dependent $T_c$ of $A_3MH_9$ hydrides, calculated using Allen-Dynes (AD) equation[55] with $\mu^* = 0.1$.

The remarkable stability and superior superconductivity of $Li_3IrH_9$ have inspired our comprehensive exploration of isostructural compounds within the $A_3MH_9$ family (A = alkali/alkaline earth metals; M = transition metals) (see Figure 3). We performed high-throughput calculations on 291 systems using our independently developed software package[56], identifying 36 dynamically stable structures at 50 GPa. Among these, 11 hydrides with $T_c$ exceeding 40 K were selected for further investigation (Figure 3). We systematically evaluated the dynamic stability threshold pressure and thermodynamic stability of these 11 hydrides. $Li_3IrH_9$ stands out as thermodynamically stable, while the remaining isotypic compounds (e.g., $Li_3RhH_9$, $Li_3CoH_9$, $Mg_3IrH_9$) are thermodynamically metastable yet lie within sufficient proximity to the convex hull. Notably, $Li_3RhH_9$, $Li_3CoH_9$ and $Mg_3RhH_9$ remain dynamically stable at pressures of 20, 10 and 25 GPa, respectively, with high $T_c$ values of 124 K, 80 K and 92 K. Additionally, $Ca_3IrH_9$ is dynamically stable under ambient pressure and achieves a high $T_c$ of 70 K.

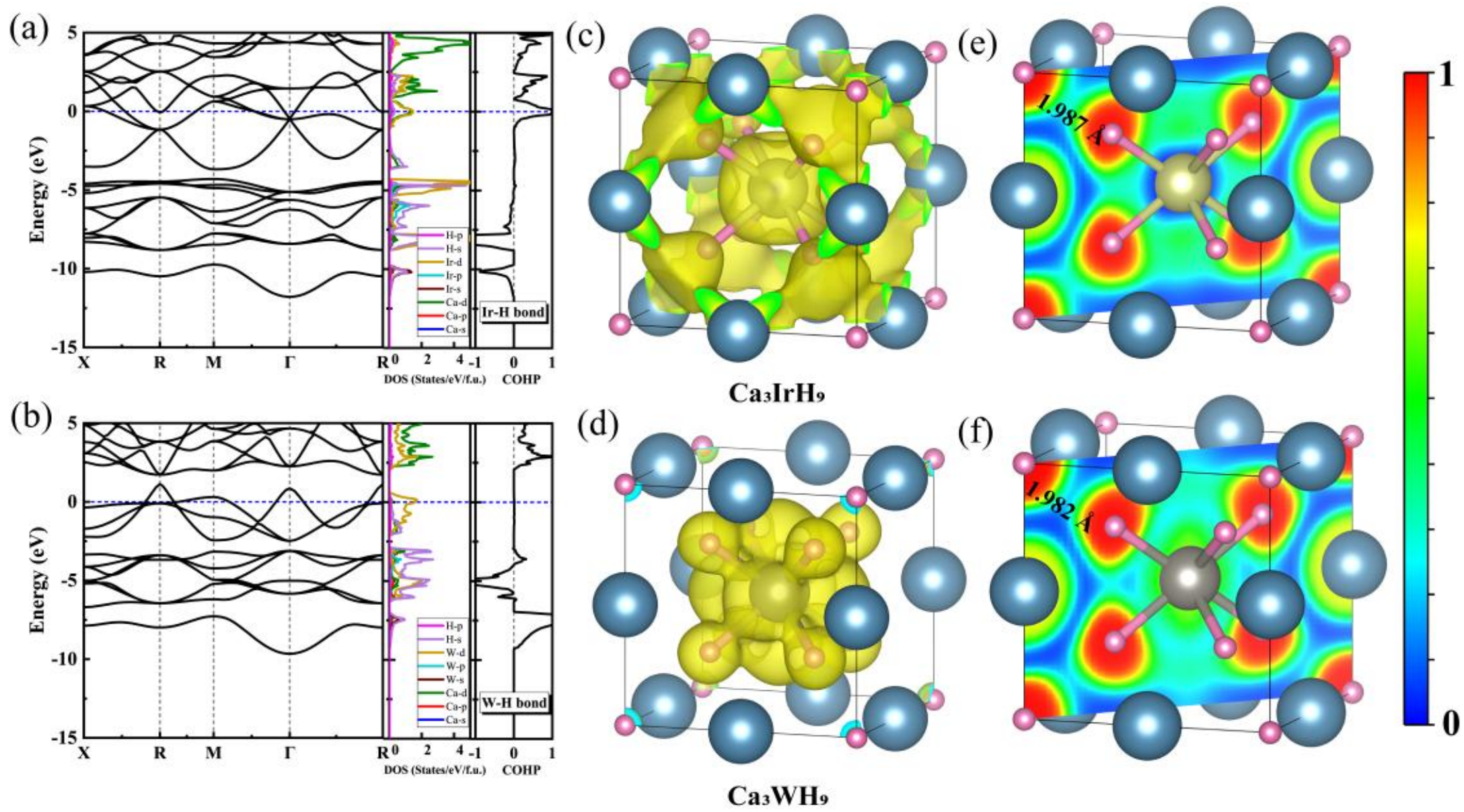

**Figure 4**. Electronic structure and bonding analysis of $Ca_3IrH_9$ and $Ca_3WH_9$ at 50 GPa. (a-b) Band structures, DOS and COHP; (c-d) Partial charge density near the Fermi level ($E_f$-2.5 eV to $E_f$); (e-f) ELF.

The ionic radii, valence electrons, and electronegativity of the elements occupying the A and M sites in the $A_3MH_9$ compounds closely influence the electronic structure near the Fermi level, thereby governing the superconducting behavior. To achieve high-$T_c$ superconductivity, A-site spectator cations preferentially favor Li and Mg due to their strong electropositive and absence of low-lying *d* orbitals, which prevent back-donation of antibonding electronic states from the [$MH_8$] anions into *d*-bands, thereby preserving hydrogen-dominated electronic states at the Fermi level[50]. M-site selection favors transition metals with electronegativities close to hydrogen (e.g., Ir and Rh), which facilitates strong orbital hybridization with hydrogen. This hybridization generates delocalized antibonding states near the Fermi level, while requiring the avoidance of excessive contributions from M-*d* states at the Fermi energy to realize hydrogen-dominated superconductivity. For example, significant hybridization between the frontier Ca-*d* and Ir-*d* orbitals in $Ca_3IrH_9$ reduces their energies relative to the Fermi level (Figure 4). Although the Ir-H bonds exhibit antibonding characteristics at the Fermi level, the symmetry matching ($A_{1g}$, $T_{2g}$) between Ca *d*-orbitals and H *s*-orbitals creates a channel for back-donation of antibonding electrons from the $[IrH_8]^{5-}$ to low-lying Ca *d* orbitals. This significantly weakens the contribution of hydrogen to the electron density of states at the Fermi level. Partial charge density near the Fermi level further confirms significant $H^{8g}$-Ca orbital hybridization, which constitutes the key basis for electron back-donation channel. Additionally, compared to $Li_3IrH_9$, the larger ionic radius of $Ca^{2+}$ induces lattice expansion, resulting in a longer $H^{1a}$-$H^{8g}$ distance (1.99 Å). The calculated ELF values between $H^{1a}$ and $H^{8g}$ were reduced, confirming the weakened coupling between them. These factors collectively lead to the significantly lower $T_c$ of $Ca_3IrH_9$. Furthermore, in $Ca_3WH_9$, the W ion is in a low oxidation state, with its frontier *d*-orbital energy levels being relatively high, and W-H bonding states occupying lower energy levels. The absence of electrons in antibonding orbitals hinders back-donation to Ca-*d* orbitals, as well as orbital overlap between $H^{1a}$ and $H^{8g}$ atoms under pressure. Calculated ELF reveals extremely weak electron localization between $H^{1a}$ and $H^{8g}$, with ionic interactions dominating the lattice. Consequently, hydrogen contributes minimally to the DOS at the Fermi level, resulting in a $T_c$ of only 1.86 K.

**CONCLUSION**

In summary, our work not only identifies a series of thermodynamically stable hydrides in the Li-Ir-H system but also demonstrates a new mechanism for driving high-temperature superconductivity in complex transition metal hydrides. The predicted $Li_3IrH_9$ stands out with robust superconductivity above 100 K across a broad pressure range (8–150 GPa), achieving a maximum $T_c$ of 121 K at 50 GPa. This exceptional performance originates from the synergistic interaction between $[IrH_8]^{2-}$ anionic units and interstitial $H^-$ ions. Specifically, the antibonding orbital delocalization in $[IrH_8]^{2-}$ overlaps with electronic states of interstitial $H^-$, driving intrinsic metallization of the hydrogen sublattice. This bonding mechanism not only generates hydrogen-dominated electron density of states at the Fermi level but also significantly enhances electron-phonon coupling through the softening of hydrogen phonon modes. Crucially, interstitial $H^-$ ions dominate electron-phonon coupling, maintaining robust contribution even under high pressure—a feature that is completely different from other known CTMHs where interstitial $H^-$ ions fail to participate in superconductivity due to insufficient orbital hybridization. Strikingly, $Li_3IrH_9$ represents the prototype of an unprecedented hydride superconductor family with the general formula $A_3MH_9$ (A=alkali/alkaline earth metal, M=transition metal). Isostructural compounds, such as $Li_3RhH_9$, $Li_3CoH_9$, and $Mg_3RhH_9$, also exhibit superior superconductivity, with maximum $T_c$ values exceeding liquid nitrogen temperature below 50 GPa. These findings establish new paradigms for achieving high-temperature superconductivity in hydride materials under moderate pressure, particularly opening the door to obtaining robust high-temperature superconductors in complex metal hydrides, and advancing practical realization of hydride superconductors under technologically accessible conditions.

**ACKNOWLEDGEMENTS**

This work was supported by the National Key Research and Development Program of China (No. 2023YFA1406200 and No. 2022YFA1402304), the National Natural Science Foundation of China (Grants No. 12274169, No. 12122405, and No. 52072188), the Program for Science and Technology Innovation Team in Zhejiang (No. 2021R01004), and the Fundamental Research Funds for the Central Universities. Some of the calculations were performed at the High-Performance Computing Center of Jilin University and using TianHe-1(A) at the National Supercomputer Center in Tianjin.

**Conflict of interest statement. None declared**